# Neutron Scattering Studies of Spin Fluctuations in High Temperature Superconductors

T.E. Mason[*]

*University of Toronto, Department of Physics,
60 St. George St., Toronto, ON Canada, M5S 1A7 and
Oak Ridge National Laboratory, Oak Ridge, TN 37831-8218*

Neutron scattering can provide detailed information about the energy and momentum dependence of the magnetic dynamics of materials provided sufficiently large single crystals are available. This requirement has limited the number of rare earth high temperature superconducting materials that have been studied in any detail. However, improvements in crystal growth in recent years has resulted in considerable progress in our understanding of the behaviour of the magnetism of the CuO planes in both the superconducting and normal state. This review will focus primarily on the spin fluctuations in $La_{2-x}Sr_xCuO_4$ and $YBa_2Cu_3O_{7-x}$ since these are the two systems for which the most detailed results are available. Although gaps in our understanding remain, there is now a consistent picture of on the spin fluctuation spectra in both systems as well as the changes induced by the superconducting transition. For both $La_{2-x}Sr_xCuO_4$ and underdoped $YBa_2Cu_3O_{7-x}$ the normal state response is characterised by incommensurate magnetic fluctuations. The low energy excitations are suppressed by the superconducting transition with a corresponding enhancement in the response at higher energies. For $YBa_2Cu_3O_{7-x}$ the superconducting state is accompanied by the rapid development of a commensurate resonant response whose energy varies with $T_c$. In underdoped samples this resonance persists above $T_c$.

---

[*] Present Address: Spallation Neutron Source, Oak Ridge National Laboratory, Oak Ridge TN, 37831-8218, USA.

# 1. Introduction

Soon after the discovery of high temperature superconductors in 1986 it was realised that superconductivity occurred in close proximity to antiferromagnetic ordering of the spin ½ $Cu^{2+}$ moments present in the $CuO_2$ planes which typify the high $T_c$ superconductors. The dynamics of these spins, both in the undoped, insulating parent compounds and in the doped, metallic, superconductors represents an intriguing topic of study, in the former case as an example of a low dimensional quantum antiferromagnet and in the latter case as a strongly correlated metal. This chapter deals with neutron scattering studies of the spin fluctuations in the $CuO_2$ planes, as such the rare earths that form an important constituent of almost all high $T_c$ materials are viewed as somewhat passive spectators. The fact that many candidate theories of high temperature superconductivity include spin fluctuations as a central feature of the mechanism for superconductivity increases the importance of obtaining a detailed picture of the energy, momentum, temperature, and doping dependence of the magnetic dynamics of the cuprates. Furthermore, many of the anomalous normal state properties of the doped cuprates are likely related to the proximity of antiferromagnetism in the phase diagram and the manifestation of this instability in the excitations. Even in the absence of a magnetic mechanism for high temperature superconductivity the spin fluctuations are an important degree of freedom of the quasiparticles which form the superconducting condensate and therefore a probe of the nature of that transition.

Information about spin excitations may be inferred indirectly from bulk measurements such as magnetic susceptibility or heat capacity (see Chapters 7 and 12). Nuclear magnetic resonance (Chapter 14) provides a more microscopic probe of the low energy spin fluctuations and can provide some information about the momentum dependence when several difference nuclear sites are compared. Raman spectroscopy (Chapter 17) can measure the two-magnon cross-section and therefore is sensitive to the energy scale over which magnetic excitations occur and its evolution with doping or temperature. The most detailed probe of spin dynamics is inelastic neutron scattering (for a recent overview see Aeppli et al. (1997a)), the topic of the present chapter. Because of its magnetic moment the neutron is sensitive to magnetic moments in solids. It is a weakly interacting, non-perturbing probe with a well-understood cross-section that is directly proportional to the static and dynamic spin correlation functions. Elastic neutron scattering can be used to determine the arrangement of spins in an ordered magnetic state (see for example, Chapter 11 describing the 4f magnetic ordering that occurs in many of the rare earth cuprate superconductors). Inelastic neutron scattering can probe crystal field excitations (Chapter 23) as well as the collective excitations of spins, ordered or disordered. While the neutron is in many respects an ideal probe of magnetism in solids it has the drawback that neutron sources are relatively weak (even the highest flux neutron source produces fewer neutrons than a bench top Cu-Kα x-ray tube. As a result, very large single crystals (of order 1 $cm^3$ or bigger) are required to measure the weak cross section inherent in any inelastic experiment. For the high $T_c$ materials this situation is made more difficult by the small spin quantum number (½) and the large energy scale of which spin fluctuations exist (100's of meV). These limitations mean that only two systems have been studied with inelastic neutron scattering in any detail, $La_{2-x}(Sr,Ba)_xCuO_4$ (214) and $YBa_2Cu_3O_{7-x}$ (123). There have been some studies of spin fluctuations in other high $T_c$ systems (for example Pr and Nd 214, see e.g. Matsuda et al. (1992) and Ivanov et al. (1995)) however, the antiferromagnetic spin waves are not qualitatively different from those observed in La 214 and the experimental difficulties of small crystals have precluded any detailed study of superconducting samples. Neutron scattering studies of 4f antiferromagnetic ordering in the rare earth high Tc superconductors is discussed in Chapter 11. Following a brief summary of the neutron scattering cross section the 214 and 123 systems are discussed in turn, with the spin fluctuations in the superconductors as the main emphasis.

# 2. Neutron Scattering Cross Section

Because of its magnetic moment the neutron can couple to moments in solids via the dipolar force. The energies and wavelengths of thermal and cold neutrons are well matched to the energy and length scales of most condensed matter systems. In the case of high temperature superconductors the characteristic energy of the spin fluctuations is determined by the superexchange interaction between nearest neighbour copper spins. This is typically ~150 meV so a complete characterisation of the magnetic dynamics requires the use of higher energy sources such as those found at spallation neutron sources or a reactor hot source. We will briefly review the formalism that describes magnetic neutron scattering. For a detailed treatment of the neutron scattering cross-section there are some excellent texts which can serve as an introduction (Squires, (1978)) or more comprehensive exposition (Lovesey, (1984)).

The partial differential cross section for magnetic neutron scattering, which measures the probability of scattering per solid angle per unit energy, is:

$$\frac{d^2\sigma}{d\Omega dE} = \frac{k'}{k}\frac{N}{\hbar}(\gamma \cdot r_o)|f(\mathbf{Q})|^2 \sum_{\alpha\beta}(\delta_{\alpha\beta} - \hat{Q}_\alpha\hat{Q}_\beta)S^{\alpha\beta}(\mathbf{Q},\omega) \quad (1),$$

where k (k') is the incident (scattered) neutron wavevector, N is the number of moments, $\gamma r_o$ = 5.391 fm is the magnetic scattering length, f(**Q**) is the magnetic form factor (analogous to the electronic form factor appearing in the x-ray scattering cross section), **Q** is the momentum transfer, ω is the energy transfer, and the summation runs over the Cartesian directions. $S^{\alpha\beta}(\mathbf{Q},\omega)$ is the magnetic scattering function which is proportional to the space and time Fourier transform of the spin-spin correlation function.

If the incident and scattered neutron energies are the same (elastic scattering) then the correlations at infinite time are being probed and, in a magnetically ordered material, the scattering function will contain delta functions at the wavevectors corresponding to magnetic Bragg reflections. The $(\delta_{\alpha\beta} - \hat{Q}_\alpha\hat{Q}_\beta)$ term in the cross section means that neutrons probe the components of spin perpendicular to the momentum transfer, **Q**. If there is no analysis of the scattered neutron energy then (within the static approximation) the measured intensity is proportional to the Fourier transform of the instantaneous correlation function which is essentially a snapshot of the spin correlations in reciprocal space. At non-zero energy transfers the spin dynamics of the system under study are probed. In a magnetically ordered system of localised spins the elementary magnetic excitations are spin waves.

The fluctuation dissipation theorem relates the correlations to absorption, in other words the scattering function is proportional to the imaginary part of a generalised (**Q** and ω dependent) susceptibility, $\chi''(\mathbf{Q},\omega)$. In the zero frequency, zero wavevector, limit the real part of the generalised susceptibility is the usual DC susceptibility measured by magnetisation. In a metal the elementary excitations are electron-hole pairs. Since it is possible to excite an electron-hole pair by promoting a quasiparticle from below the Fermi surface to above the Fermi surface, and at the same time flipping its spin, neutrons can be used to probe the low energy excitations of a metal. The generalised susceptibility (for a non-interacting metal) is just the Lindhard susceptibility that can be calculated from the band structure.

## 3. Insulating Cuprates

### *3.1 Antiferromagnetism in $La_2CuO_4$*

The high $T_c$ superconductors are complex materials displaying, in a single sample, diverse phenomena that have been active topics of research in condensed matter physics for the last thirty years. They are oxides of copper having perovskite crystal structures, generally tetragonal at high temperatures and orthorhombic at low temperatures. The importance of magnetism in the undoped, insulating, parent compounds was confirmed quite soon after the initial discovery of superconductivity in $La_{2-x}Ba_xCuO_4$ when Vaknin et al. (1987) showed, using neutron powder diffraction, that anomalies seen in the magnetic susceptibility of $La_2CuO_{4-y}$ (Johnston et al. (1987)) were due to a transition to a three dimensional, long range, ordered, antiferromagnetic state. The $Cu^{2+}$ ions of the $CuO_2$ sheets that typify this class of materials are coupled via an antiferromagnetic superexchange interaction that is considerably stronger between nearest neighbor copper spins in the nearly square $CuO_2$ planes than perpendicular to them. This leads to effectively two dimensional magnetic behavior, the transition to long range 3D order occurs due to the weak interplane interaction which is not large enough to affect the spin dynamics very significantly.

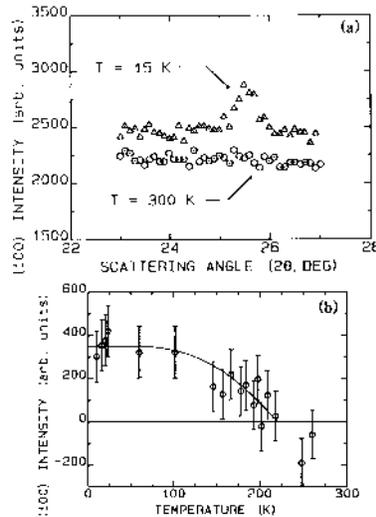

Figure 1: Antiferromagnetic ordering observed in $La_2CuO_{4-y}$ by unpolarized powder neutron diffraction. The top panel shows the (100) peak which appears below $T_N$ (=220 K), the bottom panel shows the temperature dependence of the Bragg intensity. From Vaknin et al. (1987).

Figure 1 shows the weak superlattice reflection observed below $T_N$ by Vaknin et al. (1987) in their unpolarized neutron powder diffraction measurements. The lower panel shows how this peak, which corresponds to the unit cell doubling of a simple +- arrangement of nearest neighbor spins on a square lattice, develops below $T_N$ (the actual value of $T_N$ depends on the oxygen doping, y – it is 220 K for the sample used in Figure 1). The fact that this reflection is magnetic in origin was subsequently confirmed by polarised neutron measurements (Mitsuda et al. (1987)). The ordered moment corresponds to 0.4 $\mu_B$/Cu. The ordering wavevector, (100) in orthorhombic notation or $(\pi,\pi)$ in units of the dimensionless 2D reciprocal lattice, corresponds to the Fermi surface nesting instability of the 2D square lattice from which the low temperature structure is derived.

A variety of experimental techniques have established the behavior of $La_{2-x}(Sr,Ba)_xCuO_4$ as holes are introduced into the $CuO_2$ planes by replacing La with Sr or Ba. Figure 2 summarises the results for Sr doping. Initially the effect of doping is to reduce $T_N$, it vanishes beyond x~0.03. There is a region of spin glass behavior before, for doping levels in the range 0.07-0.25, superconductivity occurs with a maximum $T_c$ of ~39 K for x=0.15. Because the $3d^9$ ionic configuration of $Cu^{2+}$ has an effective spin ½ ground state with no orbital angular momentum $La_2CuO_4$ is an excellent realisation of a 2D square lattice Heisenberg antiferromagnet – a low dimensional quantum Hamiltonian which, before the advent of high $T_c$ was not understood. There is a small spin anisotropy due to antisymmetric exchange, an indication of the extent to which this causes deviation from Heisenberg symmetry can be seen in the small canting (~0.003 rad) of the spins out of the $CuO_2$ plane (Kastner et al. (1988)). Because it is such an excellent realisation of a model quantum magnet, experiments on $La_2CuO_4$ have been crucial as a test of theories that purport to explain this system. $La_2CuO_4$ is now probably the most completely characterised model magnetic system in existence, the spin dynamics and critical scattering have been quantitatively measured and understood over an unprecedented range of energy and temperature.

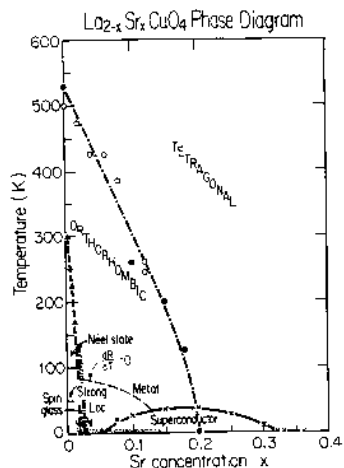

Figure 2: Phase diagram for $La_{2-x}Sr_xCuO_4$ showing the phase boundaries for the tetragonal-orthorhombic, paramagnetic-antiferromagnetic, and normal-superconducting phase transitions as a function of temperature and doping. From Keimer et al. (1992).

### 3.1.1 Spin Waves

The lamellar structure of the high $T_c$ oxides suggests that the magnetic properties of the spin ½ moments in the $CuO_2$ planes will be quasi-two-dimensional, as is the case for isostructural magnetic materials such as $K_2NiF_4$ and $K_2MnF_4$ (Birgeneau, Als-Nielsen, and Shirane, (1977)). The appropriate magnetic Hamiltonian for such a system is that of weakly coupled planes of Heisenberg spins:

$$\mathsf{H} = \sum_{ij} J_\parallel \mathbf{S}_i \cdot \mathbf{S}_j + \sum_{ij'} J_\perp \mathbf{S}_i \cdot \mathbf{S}_{j'}. \qquad (2)$$

The first term represents the coupling between nearest neighbor Cu spins in the same $CuO_2$ plane, this is the dominant interaction and neglecting the inter-layer coupling described by the second term in (2) yields the Hamiltonian of a (nearly) square lattice Heisenberg antiferromagnet. One notable feature of $La_2CuO_4$ is that it is the first good example of a Heisenberg (i.e. isotropic) spin ½ system in two dimensions studied in any detail. Unlike systems with an Ising anisotropy there is no phase transition to a long range ordered state at finite temperature expected for Heisenberg spins. The ordering below $T_N$ that occurs in $La_2CuO_4$ and other high $T_c$ compounds is driven by the weak interplane interaction that eventually leads to 3D ordering.

Well below $T_N$, in the zero temperature limit, the excitations out of the antiferromagnetically ordered ground state of equation (2) are spin waves. Neglecting interlayer coupling, conventional spin wave theory in the classical (large-S) limit predicts a dynamic susceptibility (as measured by inelastic neutron scattering) of the form:

$$\chi''(\mathbf{Q},\omega) = Z_\chi \pi g^2 \mu_B^2 S \left( \frac{1-\gamma(\mathbf{Q})}{1+\gamma(\mathbf{Q})} \right)^{1/2} \delta(\hbar\omega \pm \hbar\omega(\mathbf{Q})), \qquad (3)$$

where,

$$\hbar\omega(\mathbf{Q}) = 2J_\parallel \left[1 - \gamma^2(\mathbf{Q})\right], \qquad (4)$$

and $\gamma(\mathbf{Q}) = \cos(\pi h)\cos(\pi l)$. Equation (3) includes a parameter, $Z_\chi$, which reflects the renormalization of the magnetic response compared to the classical theory due to the fact that S is not large, and the Néel state is not a good approximation of the ground state. There is also a renormalization of the energy scale of the spin waves such that the exchange constant deduced from measurements of the spin wave dispersion relation, $J_\parallel^*$ is related to the one appearing in Equation (2) via $J_\parallel^* = Z_c J$. In the case of the S=½ square-lattice antiferromagnet Singh (1989a) and Igarashi (1992a), (1992b) have estimated $Z_c = 1.18$ and $Z_\chi = 0.51$ based on a 1/S expansion. Since neutron scattering measures only the effective dispersion it cannot be used to determine $Z_c$, however, by placing intensity measurements on an absolute scale, $Z_\chi$ can be measured directly.

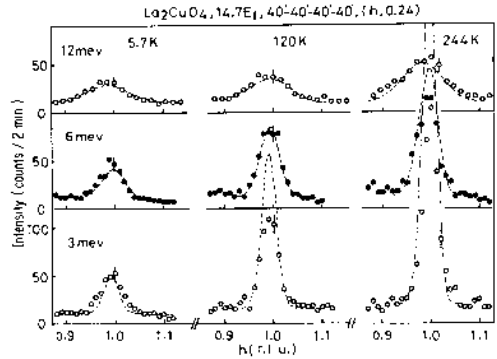

Figure 3: Thermal triple-axis constant energy transfer scans through the antiferromagnetic zone center, (100), at selected temperatures below $T_N$. The dashed lines show fits to conventional spin wave theory convoled with the spectrometer resolution, in all cases the observed peaks are unable to resolve the counter propagating spin wave modes. This type of measurement can only place a lower bound (650 meV-Å) on the spin wave velocity. From Yamada et al. (1989).

Initial attempts to measure the spin waves in $La_2CuO_4$ were frustrated by the very large value of the exchange constant $J_\parallel$, a consequence of the strong superexchange path between $Cu^{2+}$ ions via the planar O. This implies that the slope of the dispersion relation, or spin wave velocity, is very steep compared to magnetic systems typically studied using thermal neutron, triple-axis spectroscopy. For energy transfers below ~50 meV, which are easily accessed by a reactor based, thermal neutron instrument the two counter-propagating spin waves modes emerging from an antiferromagnetic reciprocal lattice point are not sufficiently well separated in momentum space to be resolved due to the finite instrumental resolution. For this reason measurements such as those shown in Figure 3 were only able to place a lower bound on the spin wave velocity (and hence exchange constant $J_\parallel$) of >400 meV-Å (Shirane et al. (1978), Endoh et al. (1988), and Yamada et al. (1989)). In order to access energy transfers large enough to separately resolve the spin waves an intense source of high energy neutrons, such as that available at from the "hot source" at the Institut Laue-Langevin, is required. A "hot source" consists of a block of graphite that is heated to a temperature much greater than the ambient temperature of the heavy water that acts a moderator for most of the neutrons. The neutrons that come to thermal equilibrium within the hot source have a Maxwellian distribution characterised by a temperature of 2000 K, as opposed to 300 K for the reactors thermalized in the heavy water. By allowing measurements to higher energy transfers the spin wave velocity can be unambiguously determined from the location of resolved spin wave peaks in constant energy scans (see Figure 4) (Aeppli et al. (1989), Hayden et al.

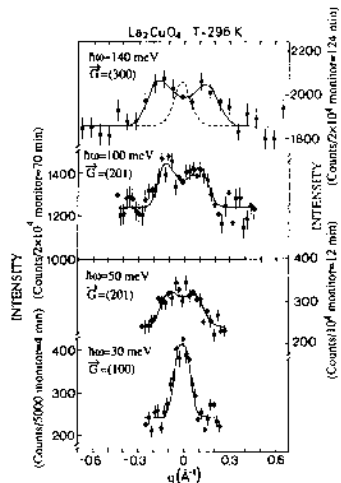

Figure 4: Using higher energy neutrons available at a "hot source" it is possible to measure the spin waves at frequencies where the two branches emerging from (100) and equivalent positions are separable, allowing a determination of the spin wave velocity. The dashed line in the upper scan shows the resolution limited response while the solid curves are the results of fits to a resolution convolved spin wave spectrum with a velocity of 850 meV-Å. From Hayden et al. (1990).

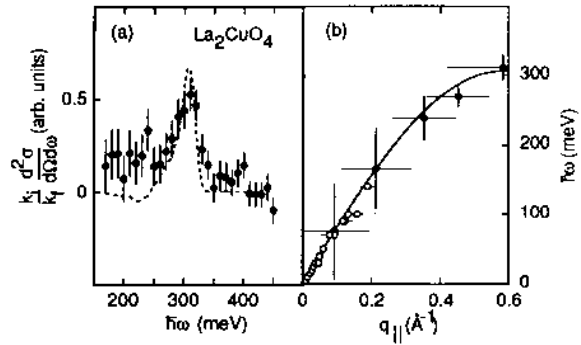

Figure 5: Spallation neutron sources have sufficient undermoderated epithermal neutrons to permit measurements with incident neutron energies up to 2 eV, allowing access the energy transfer high enough to observe the zone boundary magnon mode in $La_2CuO_4$ (panel (a)) at 312 meV. The time of flight data obtained on HET at ISIS (filled circles) combined with the triple axis measurements shown in Figure 4 (open circles) the full dispersion relation for the spin waves has been determined (panel (b)). From Hayden et al. (1991b).

(1990)). In order to fully determine the parameters of the magnetic Hamiltonian that describes the excitations in $La_2CuO_4$ it is necessary to measure spin wave excitations throughout the Brillouin zone. This can only be done using the special characteristics of a spallation neutron source that has a spectrum of undermoderated neutrons extending beyond the thermal and hot range. Figure 5 shows the zone boundary magnon for $La_2CuO_4$ measured on the HET spectrometer at ISIS along with the full dispersion relation obtained from the HET and IN1 measurements (Hayden et al. (1991b)). In addition to determining the exchange constant, $J_\parallel^* = 153 \pm 4$ meV (corresponding to a J of 135 meV), these measurements, via the ratio of the zone boundary magnon energy to the spin wave velocity (850 meV-Å), place an upper bound on the next nearest neighbor exchange constant at 9 meV. In addition, the width of observed zone boundary magnon places a lower bound on the spin wave lifetime of $9.3 \times 10^{-13}$ s.

More recent measurements of the spin waves in $La_2CuO_4$ over the full Brillouin zone have been placed on an absolute scale (Hayden et al. (1996a), Hayden et al. (1998), see also Section 4.1.2) allowing a determination of the quantum renormalization of the magnetic response. Quantitative analysis of the intensity of the spin wave response yields an estimate of $Z_\chi = 0.39 \pm 0.1$, in reasonable agreement with the calculation of Igarashi (1992a), (1992b). Taken together the inelastic neutron scattering studies if $La_2CuO_4$ provide a quantitative verification of the applicability of renormalized spin wave theory and a determination of the parameters of the magnetic Hamiltonian (Equations (2)-(4)) that describes this material.

The spin dynamics of $La_2CuO_4$ are not dramatically affected by the Néel transition, a reflection of the fact that there are quite long range 2D correlations in the paramagnetic state (see below). The three dimensional ordering is driven by the weak inter-layer coupling and has little effect on the intra-layer dynamics for energies comparable to $J^*$ beyond introducing a modest, $3.6 \pm 2.9$ meV damping (at 320 K, Hayden et al. (1990)). At low (<5 meV) energies there is a manifestation of the transition to the paramagnetic state in the form of an additional, quasielastic contribution to the cross section which has a width of $1.5 \pm 0.4$ meV, in good agreement with expectations (Grempel (1988)). This contribution is due to spin wave interactions and is also consistent with numerical simulations (Tyc et al. (1989)). For energies, $\hbar\omega \geq \hbar c \xi^{-1}$, renormalized spin wave theory gives an excellent account of the dynamics, even in the paramagnetic state.

### 3.1.2 Paramagnetic Critical Scattering

The properties of a magnetic system close to an instability such as antiferromagnetic ordering transition are characterised by divergences in various physical properties. Neutron scattering can access several of the important parameters through the energy integrated cross section, or critical scattering, which is proportional to the Fourier transform of the instantaneous spin-spin correlation function (for a detailed discussion see Collins (1989)). Of course one can numerically integrate the measured inelastic cross sections described in 3.1.1, this is not as efficient, however, as performing the energy integration experimentally by detecting scattered neutrons without energy analysis. This is accomplished on a triple axis spectrometer by operating in two-axis mode, without an analyser crystal. This will measure the desired energy integrated cross section only if the variation in momentum transfer for varying final energy can be neglected, the so-called static approximation. For a two dimensional system this can be assured by choosing the experimental geometry such that the final wavevector of the scattered neutrons is perpendicular to the planes (in this case the cross section is integrated at constant in-plane wave-vector, Birgeneau et al. (1977)). This geometry is typically employed in studies of the critical scattering in high $T_c$ cuprates. Provided the incident neutron energy is high enough to span the range of important energy transfers the instantaneous spin correlations are probed. This is an important consideration for $La_2CuO_4$ given the large exchange energy; calculations based on the cross section expected for pure $La_2CuO_4$ suggest there is not a problem in this case or for lightly doped cuprates, using conventional techniques, (Keimer et al. (1992)). The measured neutron intensity is typically a Lorentzian centered on the ordering wave-vector, (100) or $(\pi,\pi)$, with a height proportional to the staggered susceptibility and width proportional to the inverse correlation length. Figure 6 shows results obtained above $T_N$ for $La_2CuO_4$ by Keimer et al. (1992).

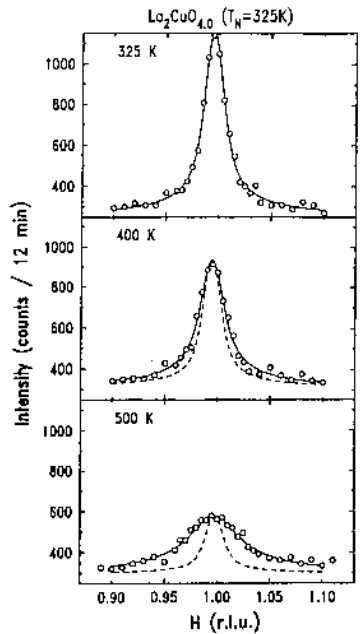

Figure 6: Energy integrating scans across the two-dimensional rod of scattering measure the Fourier transform of the instantaneous spin-spin correlation function in $La_2CuO_4$ above $T_N$. The peak width (corrected for finite instrumental resolution) is the inverse correlation length for antiferromagnetic spin fluctuations. The dashed lines show the experimental resolution function and the solid lines are the results of fits to a Lorentzian lineshape convolved with the resolution for three different temperatures. From Keimer et al. (1992).

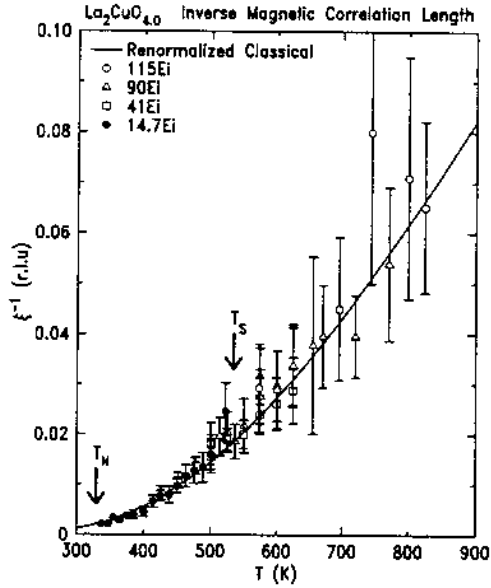

Figure 7: Inverse correlation length extracted from fits to two-axis, energy integrating scans through the antiferromagnetic position (see Figure 6). The line is the result of a fit to the renormalized classical theory described in the text. From Birgeneau et al. (1995).

Aside from a region of temperature extremely close to the 3D phase transition the antiferromagnetic correlations in $La_2CuO_4$ are essentially two dimensional in character due to the very large anisotropy in the magnetic interactions in the Hamiltonian (Equation (2)). The **Q** dependent static susceptibility of the 2D S=½ Heisenberg antiferromagnet has been treated in detail by Chakravarty et al. (1989) and Hasenfratz and Niedermayer (1991). In the regime governed by the antiferromagnetic instability (the renormalized classical region) the temperature dependence of the correlation length is determined by the exchange interaction according to the expression:

$$\xi/a = 0.493 e^{1.15J/T}\left[1 - 0.43\left(\frac{T}{J}\right) + O\left(\frac{T}{J}\right)^2\right]. \quad (5)$$

Figure 7 shows a comparison between this prediction and the measured inverse correlation length for $La_2CuO_4$ between $T_N$ and 800 K (Birgeneau et al. (1995)). The value of J extracted from the low temperature spin wave measurements is used in the comparison so there are no adjustable parameters. There is very good agreement over the range of temperatures probed with the renormalized classical predictions, perhaps to a greater extent than might be expected based on the range over which the theory is likely to be valid. Additional studies on another realisation of the spin ½, 2D Heisenberg antiferromagnet with a smaller J have covered an even wider range of T/J and found similar good agreement (Greven et al. (1994), (1996)). There is also no evidence for a cross over to a quantum critical regime as suggested based on the NMR measurements of Imai et al. (1993). Somewhat perplexingly, given the excellent agreement for the correlation length, the temperature dependence of the staggered susceptibility (peak intensity) is described by $S(0) \sim \xi^2$, not the expected $S(0) \sim \xi^2 T^2$.

## 3.2 Lightly Doped La$_{2-x}$(Sr,Ba)$_x$CuO$_4$

Doping La$_2$CuO$_4$ by replacing La with Sr or Ba introduces holes in the CuO$_2$ planes, eventually leading to an insulator-metal transition and a superconducting ground state. In addition to the changes in the charge transport the low energy magnetic properties are modified by these impurities. The Néel temperature is rapidly suppressed by introduction of these impurities due to the frustration associated with a hole that substitutes a ferromagnetic bond for an antiferromagnetic one. Beyond x~0.015 there is no longer a long range ordered antiferromagnetic state at low temperatures. The range $0.015 \leq x \leq 0.05$ is characterised by carrier localization below ~100 K and two-dimensional spin glass freezing (Harshman et al. (1988) and Sternlieb et al. (1990)). Despite the absence of long range order at these intermediate doping levels there are still strong antiferromagnetic correlations as evidence by peaks in the energy integrated neutron cross sections at the same position, $(\pi,\pi)$, as in undoped La$_2$CuO$_4$ (Keimer et al., (1992)). Qualitatively the results are similar to those described in Section 3.1.2 for the paramagnetic state of La$_2$CuO$_4$ although the inverse correlation length does not vanish, the decrease that occurs on cooling saturates at a doping dependent value (see Figure 8). The increase in $\kappa$ with temperature follows the same renormalized classical form (Eq. (5)) as the undoped material leading to a simple description where the antiferromagnetic correlations in La$_{2-x}$Sr$_x$CuO$_4$, being characterised by an inverse correlation length that varies as $\kappa(x,T) = \kappa(x,0)+\kappa(0,T)$ with $\kappa(x,0)$ varying as $(a/\sqrt{x})^{-1}$ and $\kappa(0,T)$ following the Hasenfratz and Niedermayer (1991) expression. The form of the zero temperature inverse correlation length implies the growth of correlations is limited by the mean spacing between impurities.

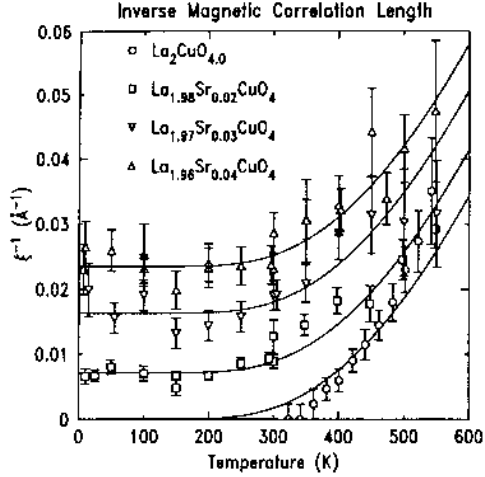

Figure 8: Inverse magnetic correlation length for antiferromagnetic spin fluctuations in lightly doped La$_{2-x}$Sr$_x$CuO$_4$. The lines are based on a simple model in which the zero temperature limit of $\kappa$ is governed by the doping and the temperature dependence is derived from the renormalized classical expression (Equation (5)). From Keimer et al. (1992).

The shorter characteristic length scale associated with lightly doped La$_2$CuO$_4$ is also seen in the magnetic dynamics. As the energy transfer is increased scans at constant energy remain centered on the $(\pi,\pi)$ position with a width that is always broader than the instrumental resolution (Hayden et al., (1991a)). The width, an inverse length scale characteristic of times ~$1/\omega$, increases linearly with energy transfer. This occurs much more rapidly than would been seen if the velocity of propagation were the same as spin waves in La$_2$CuO$_4$, there is a softening to approximately 60% of this value for the Ba$_{0.05}$ sample studied by Hayden et al. (1991a). The momentum dependence is not significantly modified by increasing the temperature to above the spin glass transition and beyond. The energy dependence of the local (**Q** integrated) susceptibility is sensitive to temperature at low energies. Figure 9 shows some data extracted from constant-**Q** scans (which are a good measure of the **Q** integrated cross section over the limited range of ω probed) which are well described by the form:

$$\chi''_o(\omega) = A \tan^{-1}\left(\frac{\omega}{\Gamma}\right). \qquad (6)$$

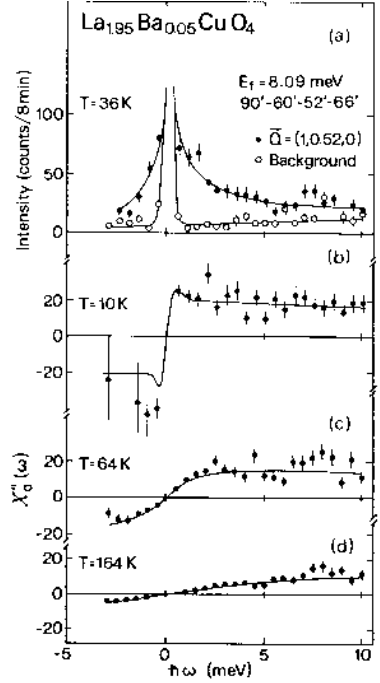

Figure 9: Constant **Q** scans at various temperatures for $La_{1.95}Sr_{0.05}CuO_4$. (a) shows the intensity at a 2D antiferromagnetic wave vector and displaced away in reciprocal space, to provide a measure of the background. (b)-(d) show the measured energy dependent response obtained by correcting the raw data for the background and thermal factor. The lines are fits to the single-site response discussed in the text. From Hayden et al. (1991a).

Above the crossover from localized to conducting transport (T~ 50 K) the lifetime $\Gamma \sim T$. This implies that the spin dynamics exhibit ω/T scaling, a result also found by Keimer et al. (1991), (1992) where the inclusion of data at low temperatures, below which logarithmic resitivity is seen, necessitates the inclusion of a higher order, $\omega/T^3$, term. This sort of scaling form for the imaginary generalised susceptibility is consistent with the marginal Fermi liquid hypothesis which has been used to account form many of the anomalous normal state properties of the cuprates (Varma et al, (1989)). The overall form taken by the temperature evolution of $\chi''(\mathbf{Q},\omega)$ indicates that the relaxation rate probed by Cu NMR is driven by the inverse lifetime, $\Gamma$, not by the correlation length, $\xi$, as derived from scaling arguments (Millis et al., (1990)).

### 3.3 Antiferromagnetism in $YBa_2Cu_3O_{6+x}$

The $YBa_2Cu_3O_{6+x}$ system is the most extensively studied of all of the families of high temperature superconductors, it has also been the topic of ongoing examination by inelastic neutron scattering. In this material the antiferromagnetically ordered parent compound is $YBa_2Cu_3O_6$, superconductivity occurs when holes are introduced into the $CuO_2$ planes by adding additional oxygen. A rough phase diagram is shown in Figure 10 which shows the suppression of $T_N$ and the onset of superconductivity which occurs for x~ 0.45.

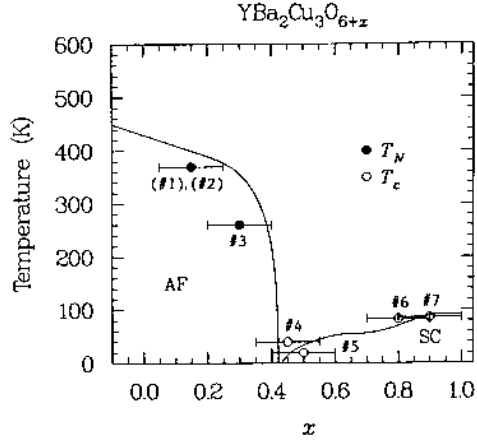

Figure 10: Phase diagram of $YBa_2Cu_3O_{6+x}$ showing the doping dependence of the antiferromagnetic and superconducting transitions. From Tranquada et al. (1989).

The magnetic dynamics of the $Cu^{2+}$ moments in insulating $YBa_2Cu_3O_{7-x}$ are governed by the same Hamiltonian as $La_2CuO_4$ (Eq. (2)). The principle difference has its origin in one of the distinguishing structural features of $YBa_2Cu_3O_{6+x}$, the $CuO_2$ bilayers. The close spacing of the $CuO_2$ planes in a bilayer (3.2 Å) give rise to an inter-layer, intra-bilayer coupling ($J_\perp \neq 0$). The inter-bilayer coupling is much smaller and can be neglected. The presence of the second term in the Hamiltonian leads to two branches in the spin wave dispersion which differs according to whether spins in adjacent layers rotate in the same ("acoustic") or different ("optic") directions (Tranquada et al. (1989)). The dynamical susceptibility for the acoustic mode is:

$$\chi''_{ac}(\mathbf{Q},\omega) = Z_\chi \pi g^2 \mu_B^2 S \left( \frac{1-\gamma(\mathbf{Q}) + J_\perp / J_\parallel}{1+\gamma(\mathbf{Q})} \right)^{1/2} \sin^2\left(\frac{\pi \Delta z l}{c}\right) \delta(\hbar\omega \pm \hbar\omega_{ac}(\mathbf{Q})) \qquad (7)$$

and for the optic mode:

$$\chi''_{op}(\mathbf{Q},\omega) = Z_\chi \pi g^2 \mu_B^2 S \left( \frac{1-\gamma(\mathbf{Q})}{1+\gamma(\mathbf{Q}) + J_\perp / J_\parallel} \right)^{1/2} \cos^2\left(\frac{\pi \Delta z l}{c}\right) \delta(\hbar\omega \pm \hbar\omega_{op}(\mathbf{Q})). \qquad (8)$$

The dispersion relations for the two modes are:

$$\hbar\omega_{ac \atop op}(\mathbf{Q}) = 2J_\parallel \{1-\gamma^2(\mathbf{Q}) + J_\perp / J_\parallel [1 \pm \gamma(\mathbf{Q})]\}^{1/2}, \qquad (9)$$

where $\gamma(\mathbf{Q}) = \frac{1}{2}[\cos(2\pi h) + \cos(2\pi k)]$ and the separation of the $CuO_2$ planes in a bilayer, $\Delta z$=3.2 Å (following conventional usage reciprocal space for $YBa_2Cu_3O_{6+x}$ is labelled according to the tetragonal notation in which the antiferromagnetic ordering occurs at wavevectors related to (½,½,l), equivalent to (π,π) in the notation of the 2D square lattice reciprocal space). The inter-planar coupling does not introduce any dispersion along the c axis, however, the $\sin^2$ and $\cos^2$ modulations can be used to distinguish between the two modes (for a more detailed treatment of the spin wave theory, including additional, smaller, terms in the Hamiltonian see Tranquada et al. (1989)).

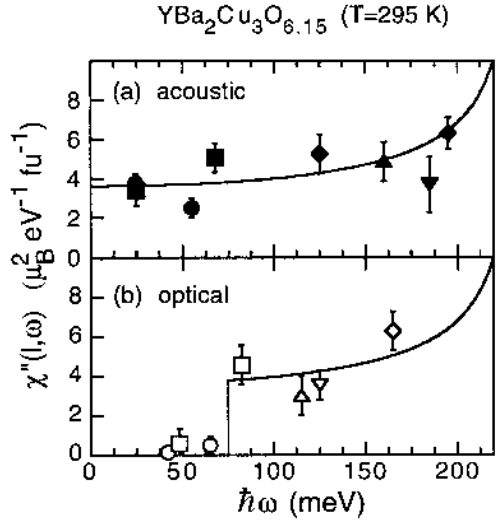

Figure 11: Energy dependence of the local susceptibility of $YBa_2Cu_3O_{6.15}$ for the (a) acoustic and (b) optic spin wave modes. The optic mode gap of 74 meV implies an inter-planar coupling of ~ 11 meV. From Hayden et al. (1998).

The optic mode gap at the magnetic zone center has a magnitude given by $\hbar\omega_g = 2\sqrt{J_\perp J_\parallel}$, the inter-layer coupling is weak but the intra-layer superexchange is comparable to the value found for $La_2CuO_4$. Due to the relatively large value of $J_\parallel$, initial studies of $YBa_2Cu_3O_{6+x}$ for antiferromagnetic compositions were not able to probe high enough frequencies to resolve the acoustic spin wave or observe the optic mode gap. Estimates of the in-plane antiferromagnetic exchange extracted from resolution corrected fits to unresolved spin waves ranged from 80 meV to 200 meV (Rossat-Mignod et al. (1988), Tranquada et al., (1989), Rossat-Mignod et al. (1991), and Shamoto et al. (1993)). The high energy spin waves in $YBa_2Cu_3O_{6.15}$ have now been resolved using the HET spectrometer (Hayden et al. (1996b), Hayden et al. (1998)) and yield a value of the exchange constant, $J_\parallel^* = 125 \pm 5$ meV, obtained from measurements of the spin waves throughout the Brillouin zone. This is consistent with the value extracted from the interpretation of two magnon Raman scattering (Singh et al. (1989b)). The same experiment also determined the quantum renormalization of the spin wave amplitude and found $Z_\chi=0.4 \pm 0.1$ in agreement with the $La_2CuO_4$ data (Hayden et al. (1996)) and the 1/S expansion (Igarashi (1992a, 1992b)). The optic mode, corresponding to out of phase motion of spins in adjacent layers of a bilayer, has also been observed by both reactor based (hot source) triple axis (Reznik et al. (1996)) and spallation source time-of-flight (Hayden et al. (1996b)) techniques. Figure 11 shows the energy dependence of the local susceptibility for values of c-axis momentum transfer chosen to probe the acoustic and optic mode response. The observed gap energy of $74 \pm 5$ meV implies $J_\perp^* = 11 \pm 2$ meV close to band theory calculations (Anderson et al. (1995)) but about twice as large as the values obtained for $Y_2Ba_4Cu_7O_{15}$ (Stern et al. (1995)) and $Bi_2Sr_2Ca_2Cu_3O_{10}$ (Statt et al. (1997)) using spin echo double resonance (a nuclear magnetic resonance technique).

As was the case for $La_2CuO_4$ renormalized spin wave theory gives a good description of the magnetic dynamics of antiferromagnetic $YBa_2Cu_3O_{6+x}$. The magnetic Hamiltonian for both compounds is the same (Equation (2)) 2D spin ½ Heisenberg antiferromagnet, however, the bilayer nature of the $YBa_2Cu_3O_{6+x}$ structure introduces an additional energy scale into the problem that is manifested in the optic spin wave modes. The exchange energy between nearest neighbor spins within a $CuO_2$ plane is somewhat smaller found in $La_2CuO_4$, possibly reflecting the larger lattice constant. The coupling between layers in a bilayer is roughly 10% of the in-plane exchange.

## 4. Superconducting $La_{2-x}Sr_xCuO_4$

Doping the single layer system $La_{2-x}Sr_xCuO_4$ to compositions where superconductivity occurs produces dramatic changes in the energy, momentum, and temperature dependence of the magnetic excitations. Although it does not possess a particularly high $T_c$ at optimal doping (~39 K) it has been extensively studied because of the relatively simplicity of its crystal structure, without bilayers and chains, and the availability of reasonable large single crystals (grown by the travelling solvent zone method) which exhibit bulk superconducting transitions with $T_c$'s close to those observed in the best powder samples.

### *4.1 Normal State Spin Fluctuations*

The normal state of the metallic cuprates are, in many respects, as anomalous as their superconductivity. The spin fluctuations at low energies in $La_{2-x}Sr_xCuO_4$ undergo a qualitative change once the doping is sufficiently large to induce metallic behavior at low temperatures. The mean impurity spacing limited, finite correlation length, antiferromagnetic response described in Section 3.2 is replaced by an incommensurate magnetic structure which, for short time scales, has a much longer characteristic length.

#### 4.1.1 Low Energy Incommensurate Response

The low energy spin dynamics of $La_{2-x}Sr_xCuO_4$ undergo a qualitative change as the doping level exceeds that required to produce metallic conductivity (and ultimately superconductivity). The broad, commensurate response of the lightly doped compositions (see Section 3.2) is replaced by an incommensurate response peaked at wavevectors displaced from the antiferromagnetic wavevector, $(\pi,\pi)$. The first indications of this were observed for samples with x=0.11 and x=0.15 (Birgeneau et al. (1989), Shirane et al. (1989), and Thurston et al. (1989)). The limitations of sample size and quality prevented determination of the location of the peak response in the planar reciprocal lattice, the temperature dependence and effect of superconductivity, and the lack of interplanar correlations. Later work by Cheong et al. (1991) showed that the magnetic response was peaked at wavevectors $(\pi,\pi) \pm \delta(\pi,0)$ and $(\pi,\pi) \pm \delta(0,\pi)$ in the notation of the 2D square reciprocal lattice (see Figure 12). The magnitude of the incommensuration, $\delta$, scales approximately as $2x$ and the intensity is rapidly suppressed as the temperature was increased to 100 K. In addition, superconductivity resulted in dramatic changes in the low energy excitations described in Section 4.2. The peaks are quite sharp in momentum space indicating that on time scales comparable to the inverse of the frequencies probed (1-15 meV) the correlations exist over distances larger than the mean spacing between impurities which limits the static correlations in lightly doped $La_{2-x}Sr_xCuO_4$. The doping dependence of the incommensuration has been studied over a wider range of compositions by Petit et al (1997) and Yamada et al. (1997). The dependence, shown in Figure 13, is similar to that of $T_c$ (see Figure 2) although there are samples with no superconducting transition that have and incommensurate response indicating it is not uniquely present in superconductors. Note that the notation used in Figure 13 and Yamada et al. (1997) is different from that of Petit et al. (1997) and Cheong et al. (1991), hence the differing scale of the incommensuration, $\delta \sim x$, the data are all consistent. The most notable feature of the behavior seen in Figure 13 is the abrupt change from commensurate to incommensurate fluctuations for x~0.06, making it coincident with the metal-insulator transition. There is also a saturation of $\delta$ beyond optimal doping (x=0.14-0.17). Full exploration of higher doping has been precluded by the increasing difficulty of growing homogeneous single crystals with x>0.2.

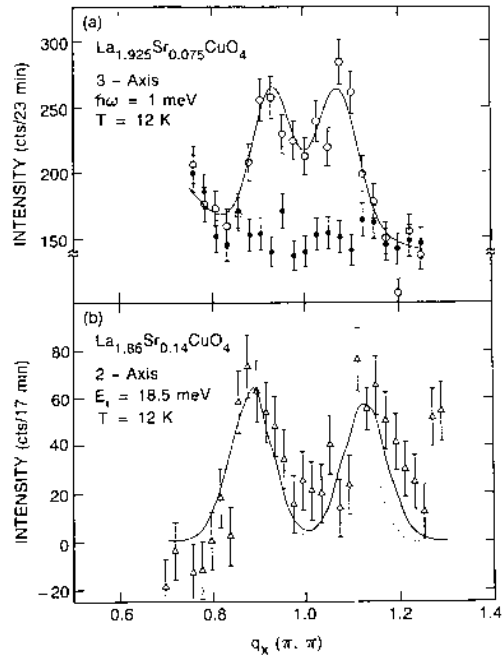

Figure 12: Low energy incommensurate magnetic response in metallic $La_{2-x}Sr_xCuO_4$ for two compositions (x=0.075 and x=0.14). The data are scans in momentum transfer through two of the four peaks that appear at $(\pi,\pi) \pm (\pi,0)$ and $(\pi,\pi) \pm (0,\pi)$. The upper panel is for an energy transfer of 1 meV, the lower is energy integrated using an incident neutron energy of 18.55 meV. From Cheong et al. (1991).

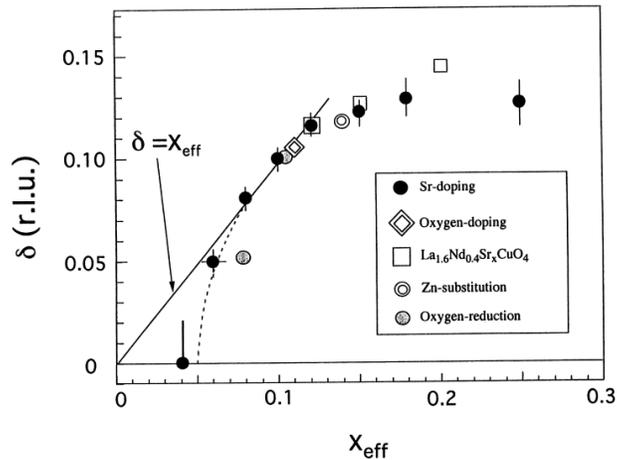

Figure 13: Incommensurability, $\delta$, of the spin fluctuations in $La_{2-x}Sr_xCuO_4$ and related systems plotted as a function of the effective doping level. The magnitude of the incommensuration scales in roughly the same way as the superconducting transition temperature with doping. From Yamada et al. (1997). Note that the notation used to describe the location of the peaks in this figure (and the original reference) differs from that used in the text by a factor of 2.

The origin of incommensurate spin fluctuations occurring as a result of the introduction of holes into the $CuO_2$ planes of $La_2CuO_4$ can be explained by a variety of models. The fact that the incommensurate response develops abruptly as the metallic compositions are encountered suggests it might be a manifestation of the Fermi surface. Incommensurate peaks in the dynamical susceptibility occur quite naturally for a 2D square lattice metal away from half filling and can be calculated on the basis of a tight binding model of the band structure (Bulut et al. (1990), Bénard et al. (1993), Littlewood et al. (1993), Si et al. (1993), and Tanamoto et al. (1994)). This description has the virtue that it leads to a simple interpretation of the effects of superconductivity on the spin fluctuations, however, there is not yet detailed agreement between the experiments and calculations in either the normal or superconducting state. An alternative view is that the incommensurate peaks are a consequence of commensurate antiferromagnetic correlations within stripes delineated by the ordering of the charged holes (Emery and Kivelson (1993) and Zaanen and Gunnarsson (1989)). Static charge and spin ordering of this type occurs in doped $La_2NiO_4$ and Nd doped $La_{2-x}Sr_xCuO_4$ (for a recent review see Tranquada (1998)). For superconducting $La_{2-x}Sr_xCuO_4$ there is no static order and attempts to observe charge peaks related to the incommensurate magnetic peaks have not been successful to date. However, the similar wavevector of the magnetic modulations in the ordered systems to the incommensurate peak positions in the superconducting compounds is suggestive. Descriptions taking the antiferromagnetic parent compound as a starting point and introducing frustration with the holes can also lead to incommensurate instabilities (Aharony et al. (1988) and Shraiman and Siggia, (1989)).

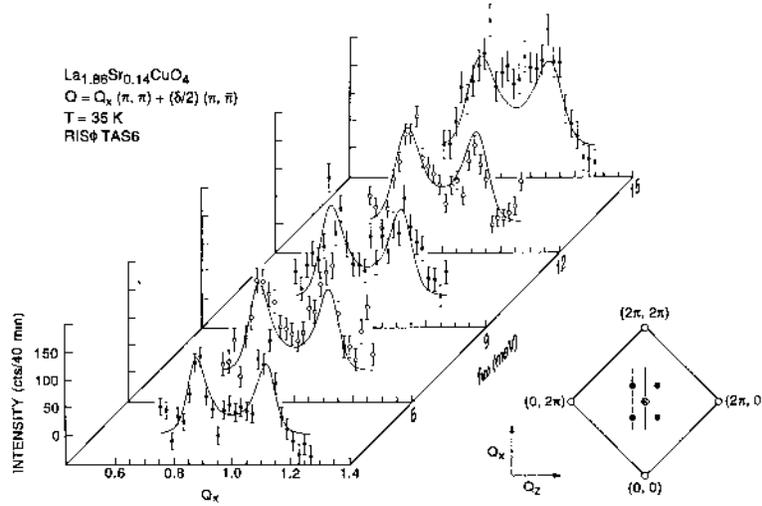

Figure 14: Energy and momentum dependence of the incommensurate spin fluctuations in the normal state of $La_{1.86}Sr_{0.14}CuO_4$. The inset in the lower left shows the scan trajectory in the 2D square lattice momentum space (dashed line). The low frequency incommensurate response broadens with increasing energy, merging into a broad commensurate feature as the energy is increased beyond ~15 meV. From Mason et al. (1992).

The evolution of the incommensurate response as the energy is increased is summarised in Figure 14 which shows scans through two of the four incommensurate peaks in $La_{1.86}Sr_{0.14}CuO_4$ at 3.5, 6, 9, 12, and 15 meV at 35 K (just above $T_c$). The incommensurate peaks are well defined at lower energies but broaden with increasing energy, the scan at 15 meV no longer displays well resolved incommensurate features. Qualitatively this (E,**Q**) dependence is very similar to that observed in Cr and Cr alloys above the spin density wave (SDW) transition (Noakes et al. (1990)), the lines in Figure 14 are the results of a fit to a functional form proposed by Sato and Maki (1974) to describe Cr (modified in this case to take account of the two dimensional nature of the incommensurate magnetism in $La_{2-x}Sr_xCuO_4$).

More recent results (Aeppli et al. (1997b)) with better signal/noise and improved counting statistics have shown that this simple, mean field, model is not sufficient to simultaneously describe the momentum and energy dependence of the spin fluctuation in the normal state of $La_{1.86}Sr_{-0.14}CuO_4$, however the general trend of the low energy peaks merging into a broad, flat topped, commensurate response at higher energies (which also occurs in the Cr systems) is captured. Beyond ~25 meV the incommensurate character of the magnetic response disappears (Petit et al., (1997)) although the very interesting regime where this incommensurate-commensurate collapse occurs has not been extensively explored due to the presence of optical phonons, making unpolarized measurements unreliable. The high energy response, which in many respects resembles that of $La_2CuO_4$ is discussed below in 4.1.2. For energies below 15 meV the spin dynamics are well described by a generalisation of the Sato and Maki (1974) form:

$$S(\mathbf{Q},\omega) = \frac{[n(\omega)+1] \cdot \chi_P''(\omega,T) \cdot \kappa^4(\omega,T)}{[\kappa^2(\omega,T) + R(\mathbf{Q})]^2} \quad (10)$$

where

$$R(\mathbf{Q}) = \frac{[(q_x - q_y)^2 - (\pi\delta)^2]^2 + [(q_x + q_y)^2 - (\pi\delta)^2]^2}{2(2a_o\pi\delta)^2}, \quad (11)$$

is a function, with the full symmetry of the reciprocal lattice and dimensions of $|\mathbf{Q}|^2$, which is positive except at zeroes, coinciding with the incommensurate peak positions. The parameters $\kappa(\omega,T)$ and $\chi''_P(\omega,T)$ describe the frequency and temperature dependent inverse length scale and peak susceptibility respectively. This form was chosen because it provided a good description of the data (when corrected for resolution effects) over a wide range of temperature (35-350 K) and energy (1-15 meV). In particular, the strong $1/q^4$ drop off in intensity away from the incommensurate peaks is required to describe the lower temperature data (a Lorentzian does not work). In addition, the shape of $R(\mathbf{Q})$ reproduces the tendency of the data to exhibit extra intensity between the incommensurate peaks, particularly as the energy (or temperature) is increased.

The results of this analysis are summarised in Figure 15 which shows the evolution of the inverse length scale $\kappa(\omega,T)$ plotted against temperature and energy added in quadrature for energies between 2.5 and 15 meV and temperatures between 35 and 350 K. All of the data follow the simple form

$$\kappa^2 = \kappa_o^2 + a_o^{-2}\left[(k_B T/E_T)^2 + (\hbar\omega/E_\omega)^2\right]^{1/Z} \quad (12)$$

where $Z=1$, $\kappa_o=0.034$ Å$^{-1}$, and $E_T=E_\omega=47$ meV ($\sim J/3$), shown as a solid line. The inset in the upper right of Figure 15 shows the variation of the low frequency limit of $\chi''_P/\omega$ with $\kappa$ which obeys a power law $\kappa^{-\delta}$ with $\delta=3$. The $(\omega,T)$ scaling, dynamical exponent $Z=1$, and $\delta=(2-\eta+Z)/Z=3$, are all consistent with a system close to a magnetic instability – the Quantum Critical Point (QCP) of a 2D insulating magnet (Sachdev and Ye (1992), Chubukov et al. (1994), and Millis et al. (1993)). The eventual saturation of $\kappa$ at a finite value $\kappa_o$ in the zero T and $\omega$ limit implies an offset from the QCP in phase space, perhaps along the lines of the schematic representation in the upper left of the Figure. The similarity of the wavevector of the dynamic incommensurate structure observed in La$_{2-x}$Sr$_x$CuO$_4$ to the static charge and spin ordering in compounds nearby in parameter space suggests stripe formation may be the driving mechanism. It has been proposed that this is an explanation for both the anomalous normal state properties of the cuprate superconductors and the underlying origin of superconductivity (Castellani et al. (1998)).

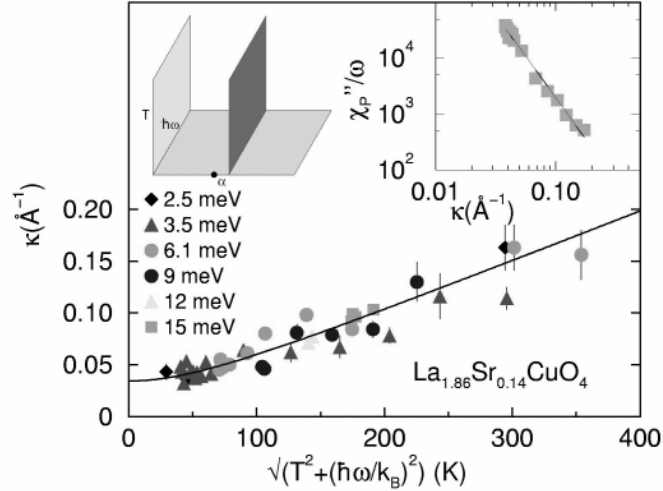

Figure 15: Temperature dependence of the inverse length scale, $\kappa(\omega,T)$, plotted at various fixed energy transfers plotted against temperature and energy added in quadrature. The solid line corresponds to a $Z=1$ quantum critical description of the data. The graph in the upper right shows the $\kappa^{-3}$ variation of the peak incommensurate response. The schematic phase diagram in the upper left depicts a 3D phase space with a quantum critical point at $\alpha$, the experiments take place in the dark vertical plane slightly displaced in phase space from the critical point. From Aeppli et al. (1997).

## 4.1.2 Comparison of High Energy Spin Dynamics with La$_2$CuO$_4$

The incommensurate fluctuations that typify the magnetic response of the metallic and superconducting compositions of La$_{2-x}$Sr$_x$CuO$_4$ are a low energy feature, merging into a broad commensurate response above ~20 meV, similar to that seen in lightly doped insulating samples. Since the magnetic response of insulating La$_{2-x}$Sr$_x$CuO$_4$ extends over energies up to 320 meV one might expect a similar energy scale to be relevant in the superconducting compositions. This issue has been explored using the HET spectrometer at ISIS by Hayden et al. (1996a, 1998). Figure 16 shows a comparison of the energy and frequency dependence of the response in La$_2$CuO$_4$ and La$_{1.86}$Sr$_{0.14}$CuO$_4$ with the intensity units normalised to allow a detailed comparison. The schematic diagrams at the top of the figure give an impression of how the spectrum is modified by doping. The response emanating from the commensurate position is broader (reflecting shorter characteristic length scale) in momentum and falls off more rapidly with increasing frequency. The difference in the two spectra is most clear in the local susceptibility, obtained by integrating the response in momentum space. Figure 17 shows this comparison for the same two samples, the doped compound has spectral weight over the same large range of energies as the antiferromagnetic insulator but there is a significant redistribution of spectral weight from high energy to a peak in the response at ~20-30 meV. In contrast the antiferromagnet the local susceptibility is peaked at the zone boundary energy (not shown) due to the van Hove singularity in the spin wave density of states. Interestingly the peak in the response in the x=0.14 sample occurs at the same energy that the incommensurate peaks merge into a broad commensurate feature.

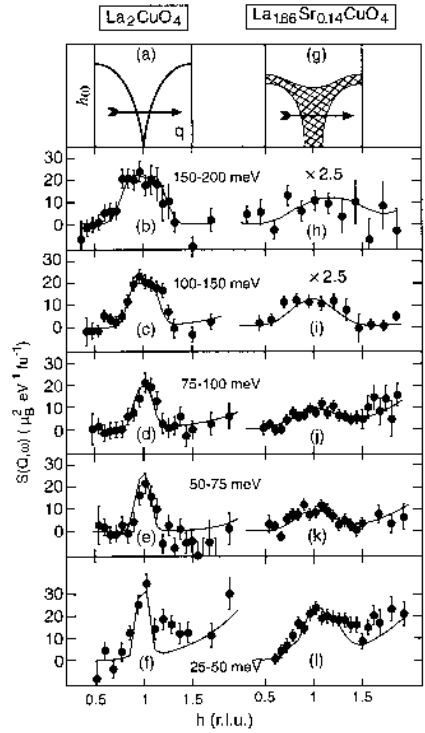

Figure 16: (a)-(f) Magnetic scattering from La$_2$CuO$_4$ and (g)-(l) from La$_{1.86}$Sr$_{0.14}$CuO$_4$ between 25 and 200 meV. All scans are in the same absolute units. The overall energy scale in the two systems is comparable but the magnetic peak is broader in the superconductor and the intensity is suppressed at higher frequencies. From Hayden et al. (1998).

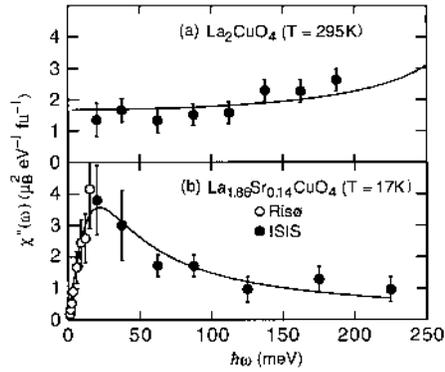

Figure 17: Local (**Q**-integrated) susceptibility for antiferromagnetic and superconducting $La_{2-x}Sr_xCuO_4$. The energies over which magnetic correlations exist is comparable for x=0 and x=0.14, however, the doped sample displays a low energy peak in the spectrum at ~20-30 meV which is not present in the insulator (where the spectrum is maximum at the zone boundary energy due to the van Hove singularity in the density of states). From Hayden et al. (1998).

## *4.2 Superconductivity and the Incommensurate Spin Fluctuations*

If the low energy incommensurate peaks in $La_{2-x}Sr_xCuO_4$ can be thought of as arising from transitions across the Fermi surface in which a quasiparticle moves from an occupied to an unoccupied state, at the same time flipping its spin, then the superconducting transition should have a dramatic effect since such a particle-hole excitation will acquire a gap of $2\Delta$. In light of this it is perhaps not surprising that the low energy spin fluctuations are suppressed by the onset of superconductivity, however, this simple physical picture ignores much of the complexity of the phenomena that was first observed by Mason et al (1992). This effect has now been studied in detail for a number of superconducting samples. The effect of superconductivity on the incommensurate response is summarised in

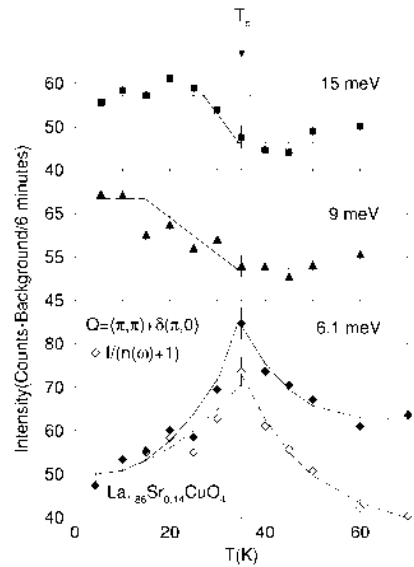

Figure 18: Temperature dependence of the incommensurate magnetic response in $La_{1.86}Sr_{0.14}CuO_4$ for 6.1, 9, and 15 meV energy transfer. For energies below 7 meV the spin fluctuations are suppressed while above that energy they are enhanced. From Mason et al. (1996).

Figure 18, which shows both the low energy suppression and the associated enhancement at somewhat higher frequencies. The crossover between suppression and enhancement may be thought of as a measure of the superconducting gap (actually twice the gap) at the wavevectors connected by the Q probed in the simple picture. For the incommensurate wave-vector this energy is 7 meV. Closer scrutiny of the data reveals subtler behavior than might be expected on the basis of the non-interacting Fermi surface particle-hole analogy. For a start, the suppression of the low energy excitations is not complete for the 6.1 meV data shown in Figure 18, the same is true down to 1.5 meV – in a clean conventional superconductor there should be no particle-hole excitations below $2\Delta$. The size of the residual low energy response is sample dependent ranging from nearly 100% of the normal state response in sample with deliberately introduced Zn impurities (Matsuda et al. (1999)) to nearly total suppression for higher quality samples (Yamada et al. (1995) suggesting an impurity induced low energy response in the superconducting state. Of course, a response at low energies does occur in a d-wave superconductor and therefore might well be expected in a high $T_c$ material. However, the calculations based on a d-wave order parameter predict a modification of the wavevector dependence of the low energy response (Zha et al. (1993)) which has not been observed (Mason et al. (1993)). Another feature of the data which has not been accounted for theoretically is the fact that the enhancement below $T_c$ for energies above 7 meV is restricted to a very narrow region of **Q** near the incommensurate peak, essentially limited by spectrometer resolution. This implies a very long coherence length, in excess of 50 Å, which is greater than any length scale present in the normal state.

## 5. Superconducting $YBa_2Cu_3O_{7-x}$

The initial measurements of the low energy spin fluctuations in $YBa_2Cu_3O_{7-x}$ for samples with oxygen content beyond the x~0.5 required for superconductivity revealed a peak, centered on the commensurate $(\pi,\pi)$ position, which is considerably broader in momentum than in insulating compositions (Tranquada et al. (1990), Chou et al. (1991), and Rossat-Mignod et al. (1988, 1991)). The evolution of the normal state commensurate response as the doping is increased beyond x=0.5 is summarized in Figure 19: the general trend of a shift in the characteristic energy of the scattering and overall decrease in intensity as the peak smears out in Q space makes the measurements increasingly difficult – to the point that for overdoped samples the $(\pi,\pi)$ response is barely observable.

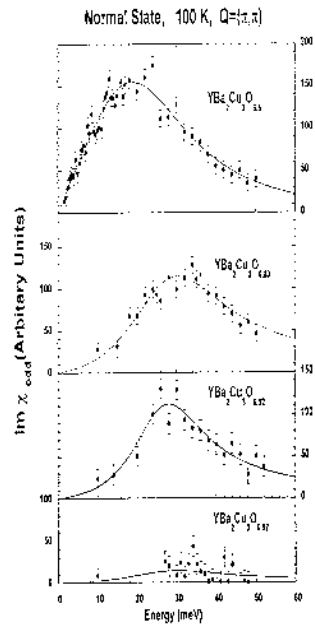

Figure 19: Low energy, commensurate [**Q**=$(\pi,\pi)$, acoustic mode] response in the normal state of four different compositions of $YBa_2Cu_3O_{7-x}$ measured at 100 K. As the doping is increased the feature at $(\pi,\pi)$ broadens and weakens, there is very little normal state response at the commensurate position for the overdoped sample. From Bourges et al. (1998).

The c-axis modulation due to the bilayer coupling which leads to an optic and acoustic mode in the spin wave cross-section for antiferromagnetically ordered samples is also present for the fluctuations in metallic samples. As was the case for the spin waves discussed in section 3.3, the low energy response is the acoustic mode, peaked at the c-axis wavevector that corresponds to a $\pi$ phase shift within a bilayer. By selecting the c-axis momentum transfer the acoustic and optic response can be probed separately, giving a determination of the optic gap, 43 meV, comparable to the spin wave gap for the optic mode (Hayden et al. (1998), Reznik et al. (1996)). Figure 20 summarizes this result for x=0.4, showing the local (**Q** integrated) susceptibility up to 200 meV (similar results have been obtained for x=0.5 (Bourges (1998), including the unexplained second peak in the acoustic response at the same energy as the peak in the optic spectrum). The most striking feature of the data shown in Figure 20 is the significant difference in overall energy variation and the absolute intensity compared to that of Figure 11 for the spin waves in a sample with x=0.85. There is a shift in spectral weight to lower frequencies in the metallic sample.

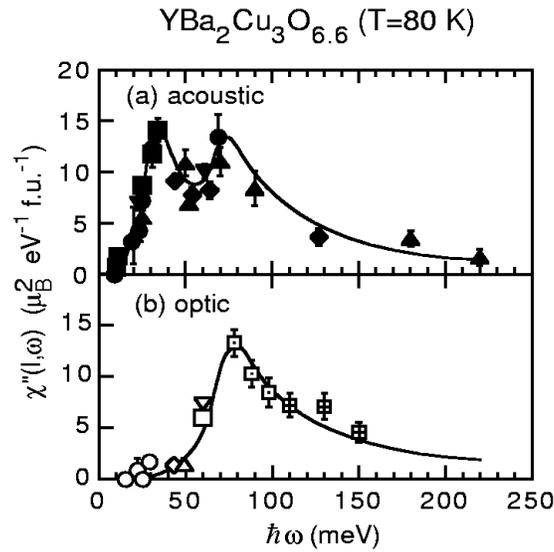

Figure 20: Normal state local (**Q** integrated) response for $YBa_2Cu_3O_{6.6}$ as a function of energy transfer. The energy gap associated with the optical mode is comparable to that observed for the spin waves in antiferromagnetic, insulating $YBa_2Cu_3O_{6.15}$. From Hayden et al. (1998).

## *5.1 Low Energy Incommensurate Response*

The fact that initial measurements of the inelastic magnetic response of superconducting compositions of $YBa_2Cu_3O_{7-x}$ found only a broad commensurate peak at low energies (in contrast to the incommensurate response found in $La_{2-x}Sr_xCuO_4$) was interpreted as arising from differences in the Fermi surface geometry of the two compounds (Si et al. (1993)). There were indications that the **Q** dependence of the cross-section for underdoped $YBa_2Cu_3O_{7-x}$ was better described in terms of unresolved incommensurate peaks, manifested as a flat topped peak at $(\pi,\pi)$ which was not well described by a single Gaussian (Sternlieb et al. (1994)). More recently Dai et al. (1998) have shown for a sample with x=0.4 (the same one as used for the data shown in Figure 20) a clear incommensurate response. Improvements to the HET spectrometer at ISIS, in the form of an array of position sensitive detectors at low angles) have allowed a complete mapping of the incommensurate structure (see Figure 21). This feature is only visible at low temperatures (T=13 K and 65 K for the data shown) and energies below ~30 meV, as was the case for $La_{2-x}Sr_xCuO_4$. Its emergence is not connected with the superconducting transition but is simply a sharpening as the temperature is lowered. The locations of the peaks are the same as for equivalently hole doped $La_{1.9}Sr_{0.1}CuO_4$ (Mook et al., (1998)).

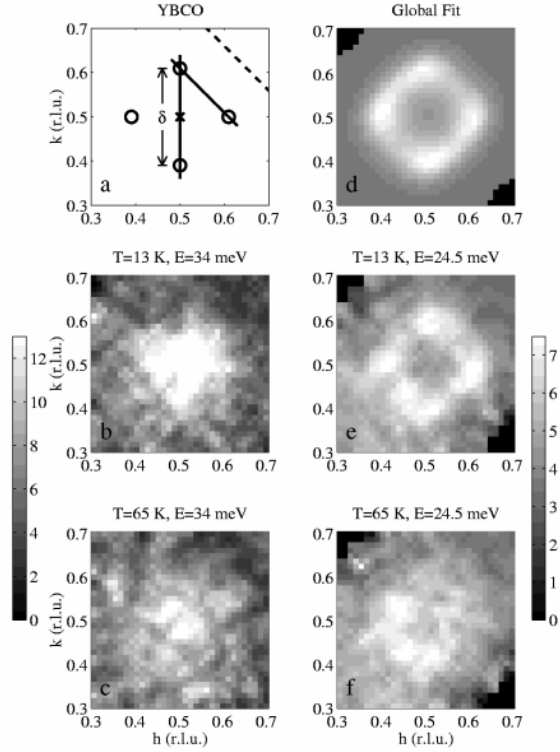

Figure 21: Images of the magnetic scattering from $YBa_2Cu_3O_{6.6}$ above and below $T_c$ at 34 and 24.5 meV in the two dimensional reciprocal space of the $CuO_2$ planes. At the lower energy (e,f) an incommensurate response, described by the model shown in d, appears at the positions noted in the schematic map, a. The resonance that appears at the $(\pi,\pi)$ position (b,c) in the superconducting state is described in 5.2. From Mook et al. (1998a).

The discovery that the low energy spin fluctuations in $YBa_2Cu_3O_{7-x}$, at least for underdoped samples, have the same incommensurate structure as $La_{2-x}Sr_xCuO_4$ suggests that this is a universal feature of high $T_c$ materials, it is also consistent with current understanding of the Fermi surface topology in both systems (see the review by Shen in this series). How the incommensuration evolves as the oxygen content is increased towards optimal doped requires the type of detailed comparative study that has been carried out for $La_{2-x}Sr_xCuO_4$ (Yamada et al. (1997)). The superconducting transition results in both a sharpening of the incommensurate peaks and an elimination of the lowest energy fluctuations; identical behavior is found in nearly optimally doped $La_{1.86}Sr_{0.14}CuO_4$ (Mason et al. (1996)).

## 5.2 The $(\pi,\pi)$ Resonance

Much of the interest in inelastic neutron scattering from high temperature superconductors has been focussed on the dramatic effects below $T_c$ often referred to as the resonance peak. Rossat-Mignod et al (1991) first observed a sharp, intense peak at $(\pi,\pi)$ that appeared below $T_c$ in optimally doped 123. The detailed energy dependence was complicated by the presence of phonons nearby in energy and momentum. The magnetic origin of the 41 meV resonance was confirmed for x=0.07 by Mook et al. (1993) using polarized neutrons (see Figure 22). This allowed the magnetic and lattice contributions to be separated and verified that the dominant feature in the magnetic response of the superconducting state was a sharp (in energy) peak at 41 meV, centered on the commensurate, $(\pi,\pi)$, position. As the oxygen content is decreased the energy of the resonance peak decreases, following the same trend as the superconducting transition temperature. For x=0.4 the resonance energy has been reduced to 34 meV, the overall variation with $T_c$ is summarized in Figure 23.

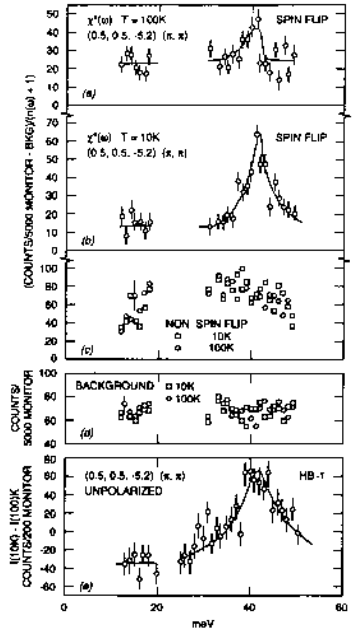

Figure 22: Polarized beam measurements of the magnetic excitations in $YBa_2Cu_3O_7$ confirming the magnetic origin of the resonance feature that appears at the $(\pi,\pi)$ position at 41 meV below $T_c$. From Mook et al. (1993).

There is a clear anomaly in the temperature dependence at the superconducting transition with an abrupt jump in intensity above the level of a weaker normal state response. The existence of a weak normal state feature was disputed (see Fong et al. (1995)) however the different measurements were not performed on samples with the same oxygen content. The general trend of the normal state response shown in Figure 19 indicates a gradual elimination of the intensity at $(\pi,\pi)$ in the normal state as the oxygen content is increased beyond optimal doping (x~0.07).

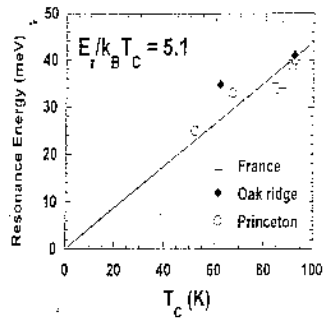

Figure 23: Variation of the $(\pi,\pi)$ resonance energy with superconducting transition temperature, $T_c$. From Bourges (1998).

For underdoped samples, where the normal state intensity is quite significant, the presence of a peak at essentially the same energy as the resonance above $T_c$ is clear (see Figure 24). The growth in the intensity at the resonance energy begins well above $T_c$ although there is a rapid increase, and an associated reduction in energy width, which occurs just below $T_c$. The size of the peak in the normal state is reduced as the oxygen content is increased and the temparture at which it becomes apparent is also reduced (this onset temperature appears to track the pseudo gap feature seen by NMR) (Dai et al. (1998)).

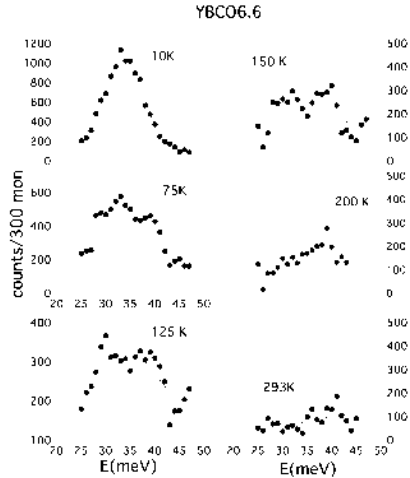

Figure 24: Temperature dependence of the 35 meV resonance in $YBa_2Cu_3O_{6.6}$ with temperature. A broadened response at $(\pi,\pi)$ persists in the normal state for underdoped compositions. From Mook et al. (1998b).

There are a variety of theoretical descriptions of the origin of the $(\pi,\pi)$ resonance ranging from a superconducting coherence effect which occurs in the framework of an itinerant electron picture (Ohashi et al. (1993), Monthoux and Scalapino (1994), Lu (1992), Bulut and Scalapino (1996), Lavagna and Stemmann (1994), Onufrieva and Rossat-Mignod (1995), and Liu et al. (1995)) to a collective mode associated with a multicomponent, SO(5), order parameter (Zhang, (1997)). In almost all description of the resonance peak it is a direct consequence of the d-wave symmetry of the order parameter. In many cases, however, the existence of a resonance above the superconducting transition poses a challenge.

## 6. Conclusions

The experimental difficulties posed by the large energy scale, weak spin ½ cross-section, and (for metallic samples) short characteristic length scale in high temperature superconductors has necessitated advances in the state of the art in inelastic neutron scattering in order to develop an overall picture of the important features. The improved capabilities of neutron instrumentation, together with significant advances in crystal growth, particularly of large, high quality crystals needed for this type of work, have led to a fairly consistent picture of spin fluctuations in the cuprates. Generally, when measurements on the same compounds (at the same doping level) exist, there is good agreement between different groups. The importance of these measurements to the field of high temperature superconductivity as a whole rely on the special characteristics of the neutron as a probe. It is the only technique which permits measurements of spin dynamics at finite momentum, with good energy and **Q** resolution.

The antiferromagnetically ordered insulators are well described by conventional spin wave theory. The destruction of long range antiferromagnetic order with doping leads to novel scaling behavior in a spectrum which is still reminiscent of the spin waves. Metallic cuprates have a low energy response that is incommensurate and qualitatively quite different from the insulators. Furthermore it is greatly affected by the superconducting transition At higher energies the changes are much less dramatic as a function of doping and temperature. All of these features are shared by both families that have been extensively studied, $La_{2-x}Sr_xCuO_4$ and $YBa_2Cu_3O_{7-x}$. The one outstanding difference is the $(\pi,\pi)$ resonance which has only been seen in $YBa_2Cu_3O_{7-x}$. A search for a commensurate resonance in $La_{2-x}Sr_xCuO_4$ is probably warranted given the otherwise similar picture which has emerged of the overall behavior with 123.

## 7. Acknowledgements

I would like to thank my colleagues for allowing me to reproduce figures, providing material prior to publication, and stimulating discussions. I would particularly like to thank my collaborators in experiments I have been involved in for their patience and guidance. I would also like to acknowledge the financial support received from the Natural Sciences and Engineering Research Council of Canada, the Canadian Institute for Advanced Research, and the Alfred P. Sloan Foundation. Work at Oak Ridge was supported by the US Department of Energy.

# Table of Figures